\documentclass{DISproc}

\begin{document}
\title{Imaging partons in exclusive scattering processes}

\author{{\slshape Markus Diehl}\\[1ex]
DESY, Notkestra{\ss}e 85, 22607 Hamburg, Germany
}

\contribID{98}

\doi  

\newcommand{\jpsi}{J\mskip -2mu/\mskip -0.5mu\Psi}

\maketitle

\begin{abstract}
  The spatial distribution of partons in the proton can be probed in
  suitable exclusive scattering processes.  I report on recent performance
  estimates for parton imaging at a proposed Electron-Ion Collider.
\end{abstract}


Deeply virtual Compton scattering (DVCS) and exclusive meson production in
lepton-proton collisions offer unique possibilities for determining the
spatial distribution of quarks, antiquarks and gluons as a function of
their longitudinal momentum inside the proton.  Such ``tomographic
images'' of the proton can provide insight into key aspects of QCD
dynamics, such as the interplay between sea quarks and gluons, the
relation between sea and valence quarks, and the orbital angular momentum
carried by these partons.

To obtain such images one needs to measure the transverse momentum
transfer $\boldsymbol{\Delta}$ to the proton.  A Fourier transform then
gives the distribution in impact parameter $\boldsymbol{b}$, which is the
position of the struck parton in the plane transverse to the proton
direction of movement.  In the scattering amplitude the longitudinal
momentum fraction $x$ of the parton is integrated over, with typical
values around $\frac{1}{2} x_B$ for DVCS and $\frac{1}{2} x_V =
\frac{1}{2} x_B\, (1 + M^2 /Q^2)$ for the production of a meson with mass
$M$.  The reconstruction of a joint density in $x$ and $\boldsymbol{b}$
can be envisaged in the framework of generalized parton distributions
(GPDs) by making use of their evolution in the resolution scale, which is
given by $Q^2$ in DVCS and by $Q^2 + M^2$ in meson production.  This
requires precise data in a very wide range of $x_B$ and $Q^2$.

Measurements from HERA suggest that at $x$ around $10^{-3}$ gluons have a
more narrow impact parameter distribution than sea quarks.  Information
about the spatial distribution of valence quarks can be inferred from
electromagnetic form factors and from lattice calculations, and a direct
study of quarks with large momentum fraction $x$ will be possible with the
12 GeV upgrade at Jefferson Lab.  The planned DVCS measurement by COMPASS
will give us a first glimpse into the $x$ region between $10^{-1}$ and
$10^{-2}$.  To realize the full physics potential of parton imaging will,
however, require a new facility.  Here I report on a study of parton
imaging at a proposed electron-ion collider (EIC) \cite{Boer:2011fh}.  For
complementary information see \cite{Fazio:DIS,Muller:DIS}.


Pseudo-data for $ep\to ep\gamma$ have been generated according to a GPD
model that reproduces the existing DVCS measurements of H1 and ZEUS.
Technically, the unpolarized distributions $H$ for gluons and for sea
quarks are modeled with two SO(3) partial waves each, as described in
\cite{Muller:DIS} and in Sec.~3 of \cite{Kumericki:2009uq}.  The $t$
dependence of the distributions is $\exp\bigl[ \,t\, (\frac{B}{2} +
\alpha' \log\frac{1}{x}) \bigr]$, with different slopes $B$ for gluons and
sea quarks and with a small shrinkage parameter $\alpha'$ as suggested by
HERA measurements.  In DVCS, the invariant momentum transfer $t$ to the
proton and its transverse component are related by $-t =
(\boldsymbol{\Delta}^2 + x_B^2\, m_p^2) /(1-x_B^{})$, where $m_p$ is the
proton mass.  In the simulation \cite{Fazio:DIS} acceptance cuts for the
final-state electron, photon and proton have been imposed, and the data
have been smeared for the expected resolution in $x_B$, $Q^2$ and $t$.  An
assumed $5\%$ uncorrelated systematic error has been added in quadrature
to the statistical error in the cross section.  The cross section
$d\sigma/dt$ for DVCS ($\gamma^* p\to \gamma p$) is obtained after
subtraction of the cross section for the Bethe-Heitler process, with an
uncertainty of $3\%$ taken on the latter.  Beam energies are $E_e = 20
\operatorname{GeV}$ and $E_p = 250 \operatorname{GeV}$.  Statistical
errors are for an integrated luminosity of $10 \operatorname{fb}^{-1}$ for
the $|t|$ range from $0.03\operatorname{GeV}^2$ to $1 \operatorname{GeV}^2$
and for $100 \operatorname{fb}^{-1}$ for $|t|$ above
$1\operatorname{GeV}^2$.  Here and in the following the uncertainty in the
overall 
luminosity is not included in the errors, because it does not affect the
form of the spectra which are at the center of our interest.

\begin{figure}
  \centering
  \includegraphics[width=0.49\textwidth]{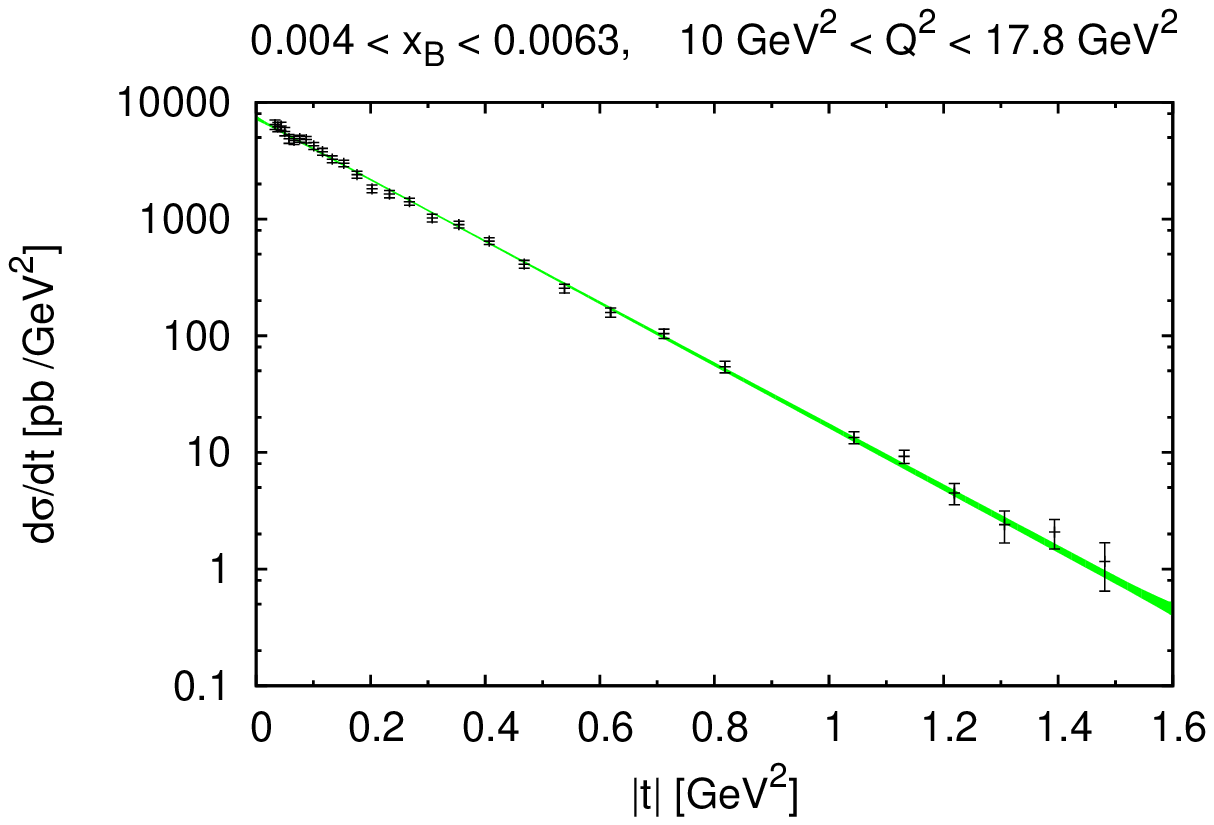}
  \includegraphics[width=0.5\textwidth]{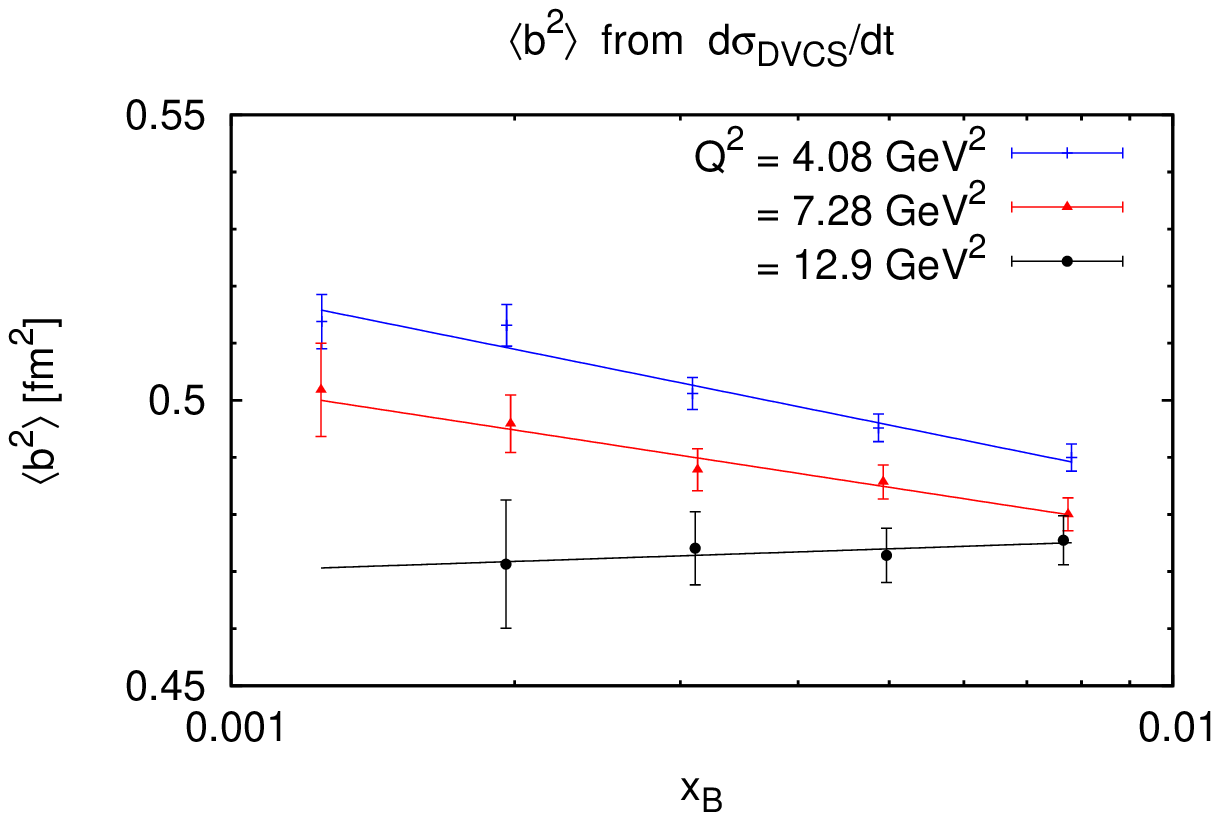}
\caption{Left: simulated DVCS cross section at EIC for a bin in $x_B$
    and $Q^2$.  Right: average squared impact parameter obtained from
    $d\sigma/dt_{\text{DVCS}}$ for different bins in $x_B$ and $Q^2$.}
  \label{Fig:sigma-dvcs}
\end{figure}

\begin{wrapfigure}{L}{0.49\textwidth}
  \centering
  \includegraphics[width=0.49\textwidth]{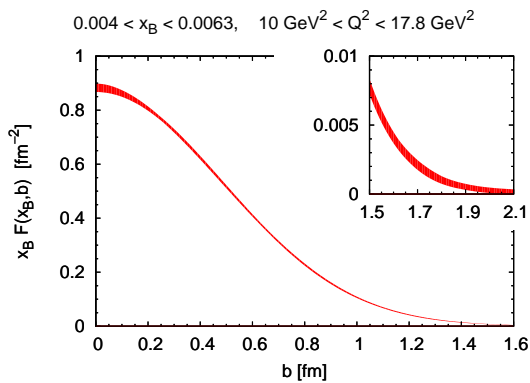}
  \caption{Impact parameter distribution in DVCS obtained from the left
    panel of Fig.~\protect\ref{Fig:sigma-dvcs}.  The band reflects the
    uncertainty in fitting $d\sigma/dt$ and in extrapolating it to the
    unmeasured $t$ region as specified in Sec.~3.6.1 of
    \protect\cite{Boer:2011fh}.}
  \label{Fig:bspace-dvcs}
\end{wrapfigure}

The left panel of Fig.~\ref{Fig:sigma-dvcs} shows the resulting $t$
spectrum for DVCS in a bin of $x_B$ and $Q^2$, together with an
exponential fit.  From the $t$ slope one obtains the average squared
impact parameter $\langle \boldsymbol{b}^2 \rangle$ for the particular
combination of quarks, antiquarks and gluons ``seen'' in DVCS.  The right
panel of Fig.~\ref{Fig:sigma-dvcs} shows that with the expected accuracy
one can resolve the separate dependence of $\langle \boldsymbol{b}^2
\rangle$ on $Q^2$ and $x_B$, which has never been possible so far.  In the
model used for generating the data, both dependences are small logarithmic
effects, which reflect both perturbative and non-perturbative dynamics of
sea quarks and gluons in the proton.
From $d\sigma/dt$ one obtains the Compton scattering amplitude
$|\mathcal{A}_{\gamma^* p\to \gamma p}|$, whose Fourier transform w.r.t.\
$\boldsymbol{\Delta}$ gives an impact parameter distribution $F(x_B,
\boldsymbol{b})$ of partons with momentum fraction of order $\frac{1}{2}
x_B$.  Figure~\ref{Fig:bspace-dvcs} shows that precise imaging is possible
in a wide range of $b$.  Of particular interest at large $b$ is the
prediction of an exponential falloff with an $x$ dependent slope of
typical size $1/(2 m_\pi) \approx 0.7 \operatorname{fm}$
\cite{Strikman:2003gz}.  Due to partons inside virtual pions, which
because of their small mass can fluctuate to large distances, this
predicted behavior is a consequence of chiral symmetry breaking in QCD and
awaits experimental verification.

\begin{figure}[htb]
  \centering
  \includegraphics[width=0.49\textwidth]{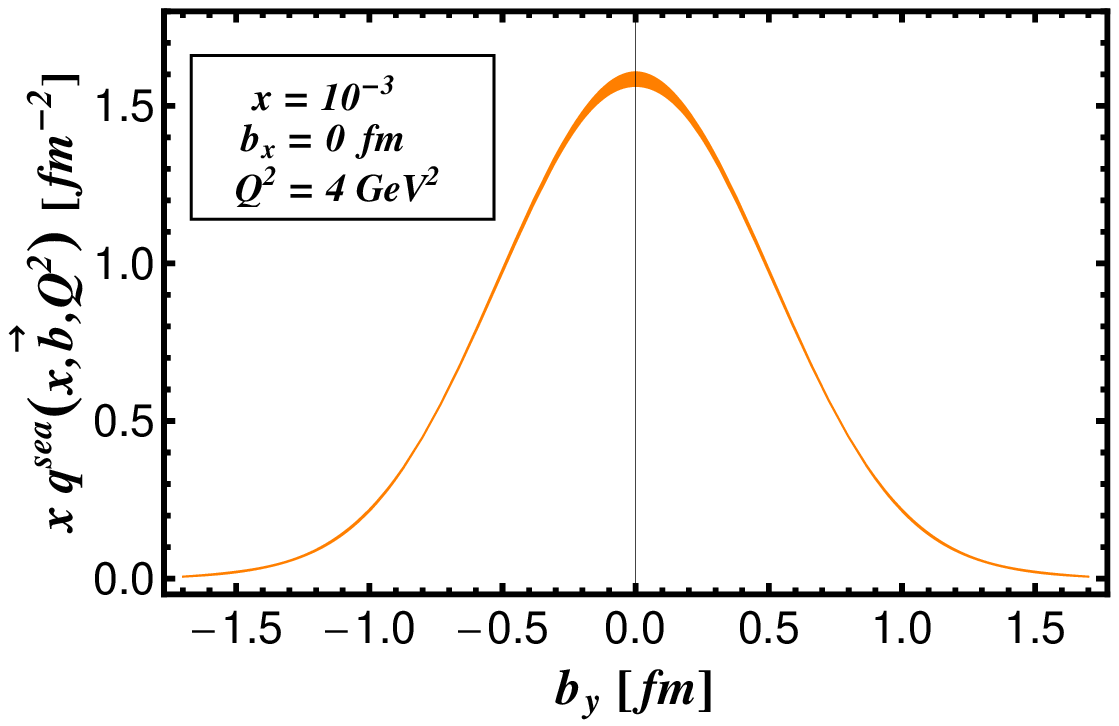}
  \includegraphics[width=0.49\textwidth]{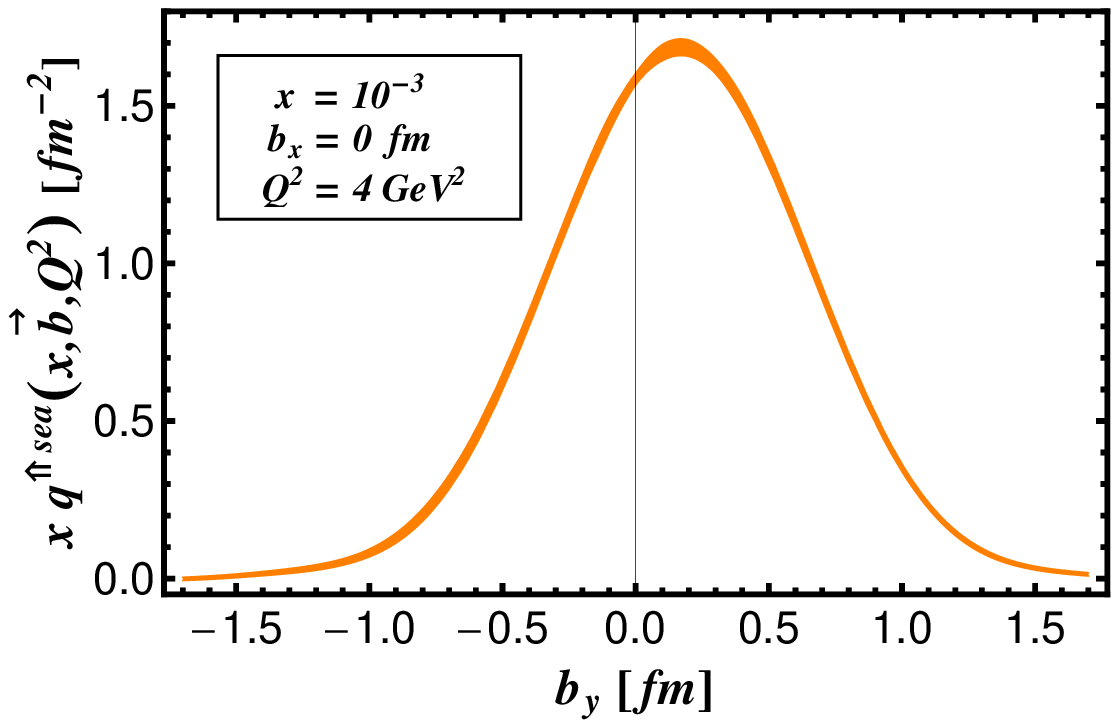}
  \caption{Impact parameter densities of unpolarized sea quarks in an
    unpolarized proton (left) and in a proton polarized along the $x$ axis
    (right), obtained from a fit to pseudo-data for the DVCS cross section
    and the transverse proton spin asymmetry
    \protect\cite{Fazio:DIS,Muller:DIS}.}
  \label{Fig:sea-gpds}
\end{figure}

\begin{wrapfigure}{R}{0.49\textwidth}
  \centering
  \includegraphics[width=0.49\textwidth]{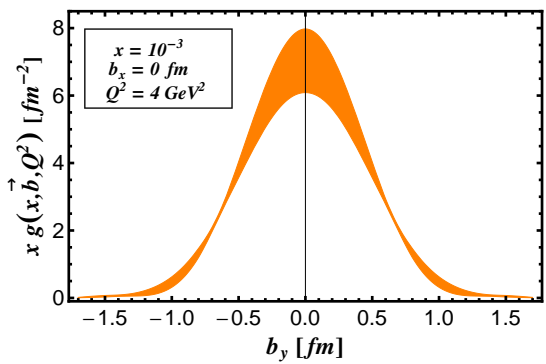}
  \caption{As Fig.~\protect\ref{Fig:sea-gpds}, but for unpolarized gluons
    in an unpolarized proton.}
  \label{Fig:gluon-gpds}
\end{wrapfigure}

With transverse proton polarization one gains access to distributions $E$,
which carry characteristic information about orbital angular momentum of
partons in the proton.  Transformed to $b$ space, these distributions
describe how the impact parameter distribution of partons is shifted
sideways in a transversely polarized proton.  Pseudo-data for the DVCS
cross section and for the transverse proton spin asymmetry have been
generated \cite{Fazio:DIS} assuming a model of $E$ for sea quarks and for
gluons of the same type as the model of $H$ described above.  Parameter
are chosen to satisfy the positivity requirements on $E$.  The simulation
is for $E_e = 20 \operatorname{GeV}$ and $E_p = 250 \operatorname{GeV}$,
with errors for an integrated luminosity of $100 \operatorname{fb}^{-1}$
and for $80\%$ transverse proton polarization measured with $5\%$
accuracy.  Systematic uncertainties of $5\%$ are added in quadrature.
As shown in Fig.~\ref{Fig:sea-gpds}, a fit of the generated data for the
DVCS cross section and the transverse proton spin asymmetry allows the
simultaneous extraction of the GPDs $H$ and $E$ for sea quarks.  The fitted
data covers the kinematic range $3.2 \operatorname{GeV}^2 < Q^2 < 17.8
\operatorname{GeV}^2$, $10^{-4} < x_B < 10^{-2}$ and $0.03
\operatorname{GeV}^2 < |t| < 1.5 \operatorname{GeV}^2$.  Thanks to the
effect of logarithmic scaling violations, even the extraction of $H$ for
gluons is possible in this fit, as shown in Fig.~\ref{Fig:gluon-gpds},
whereas the errors on $E$ for gluons are very large.

\begin{figure}[htb]
  \centering
  \includegraphics[width=0.49\textwidth]{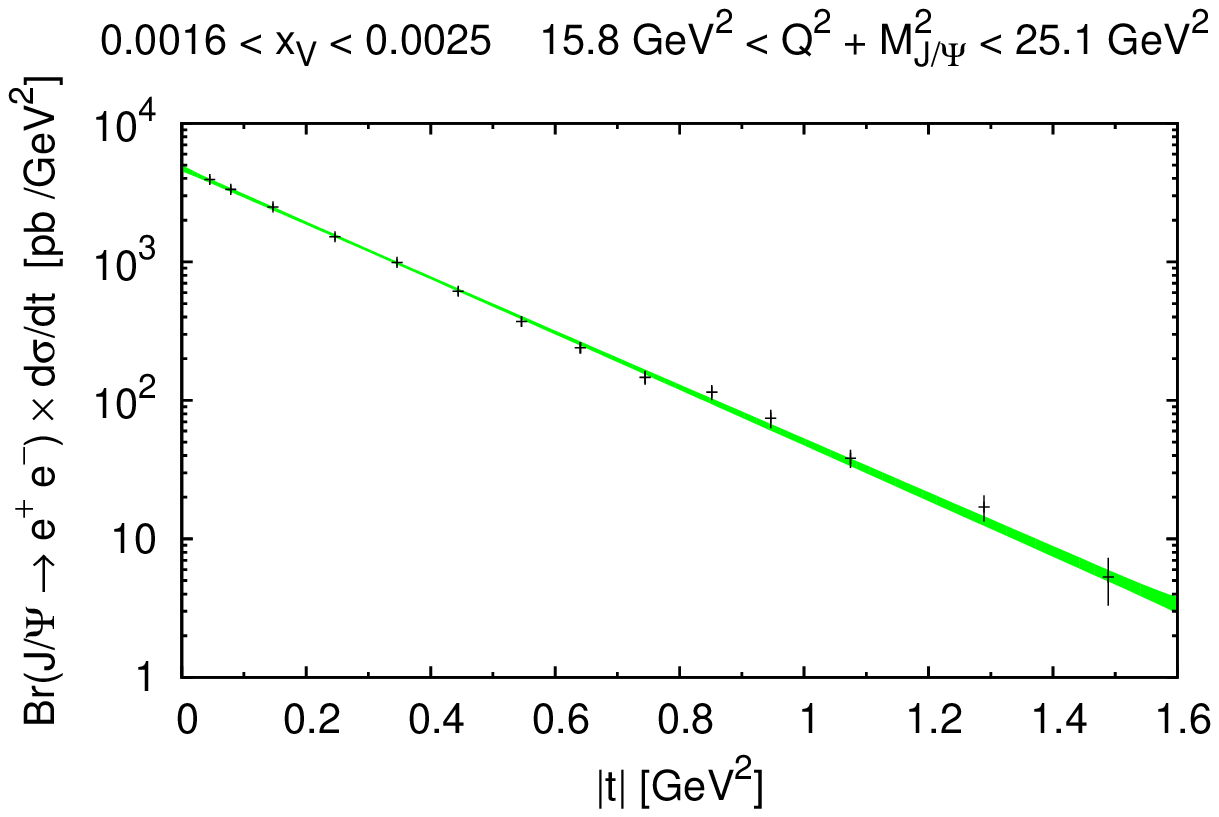}
  \includegraphics[width=0.49\textwidth]{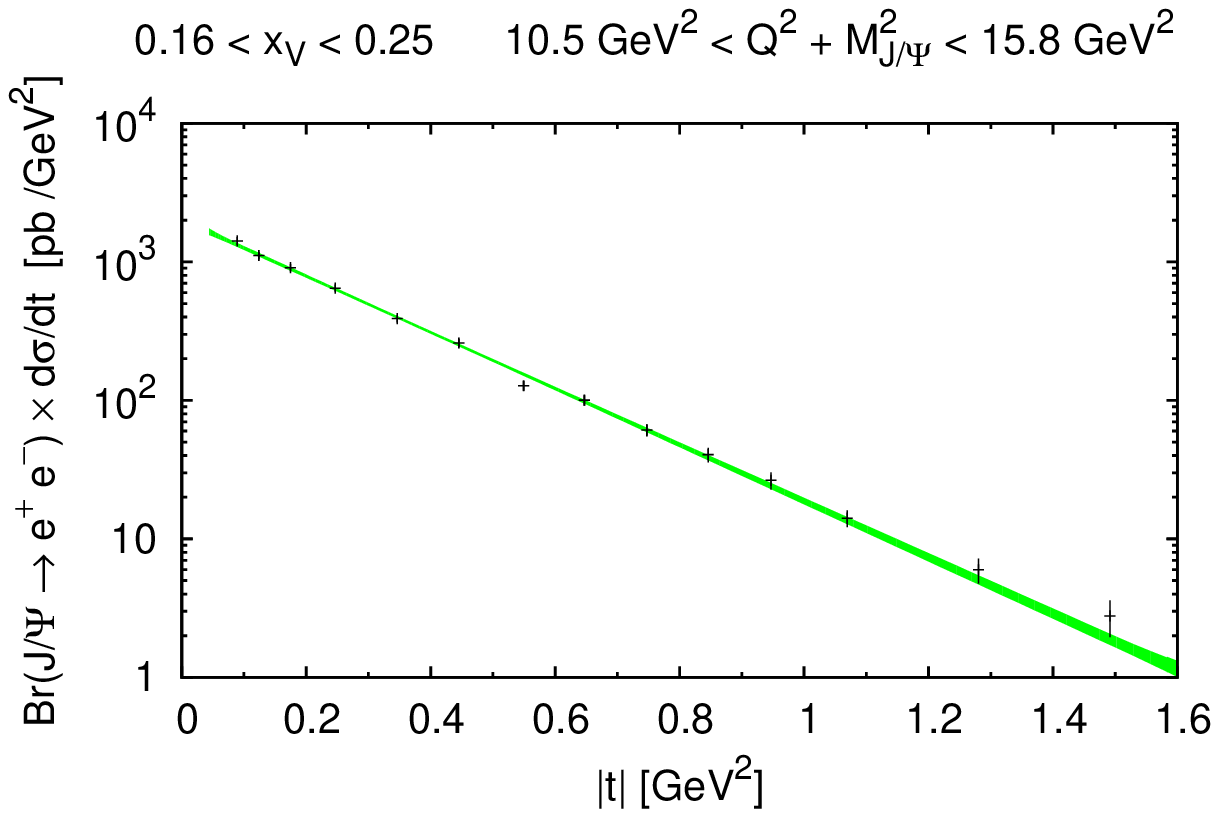} \\
  \includegraphics[width=0.49\textwidth]{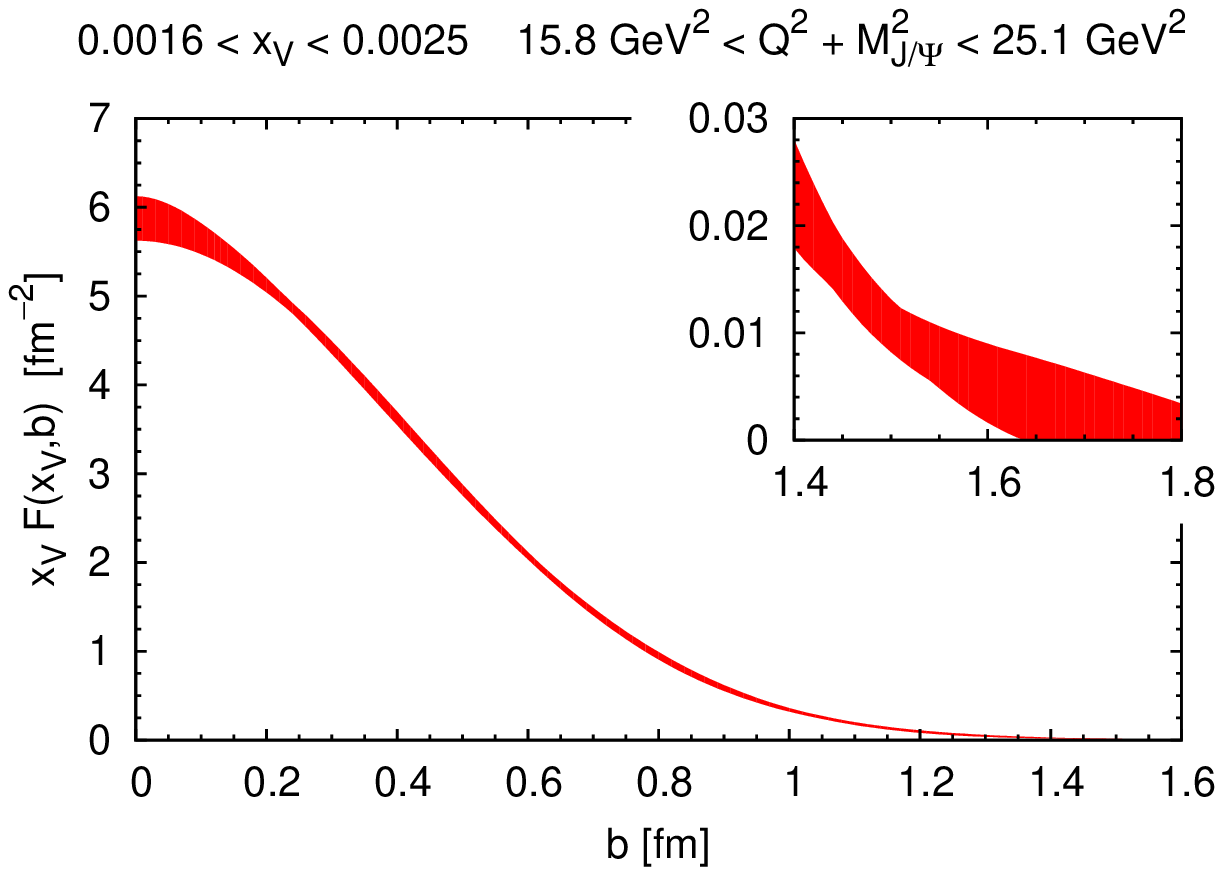}
  \includegraphics[width=0.49\textwidth]{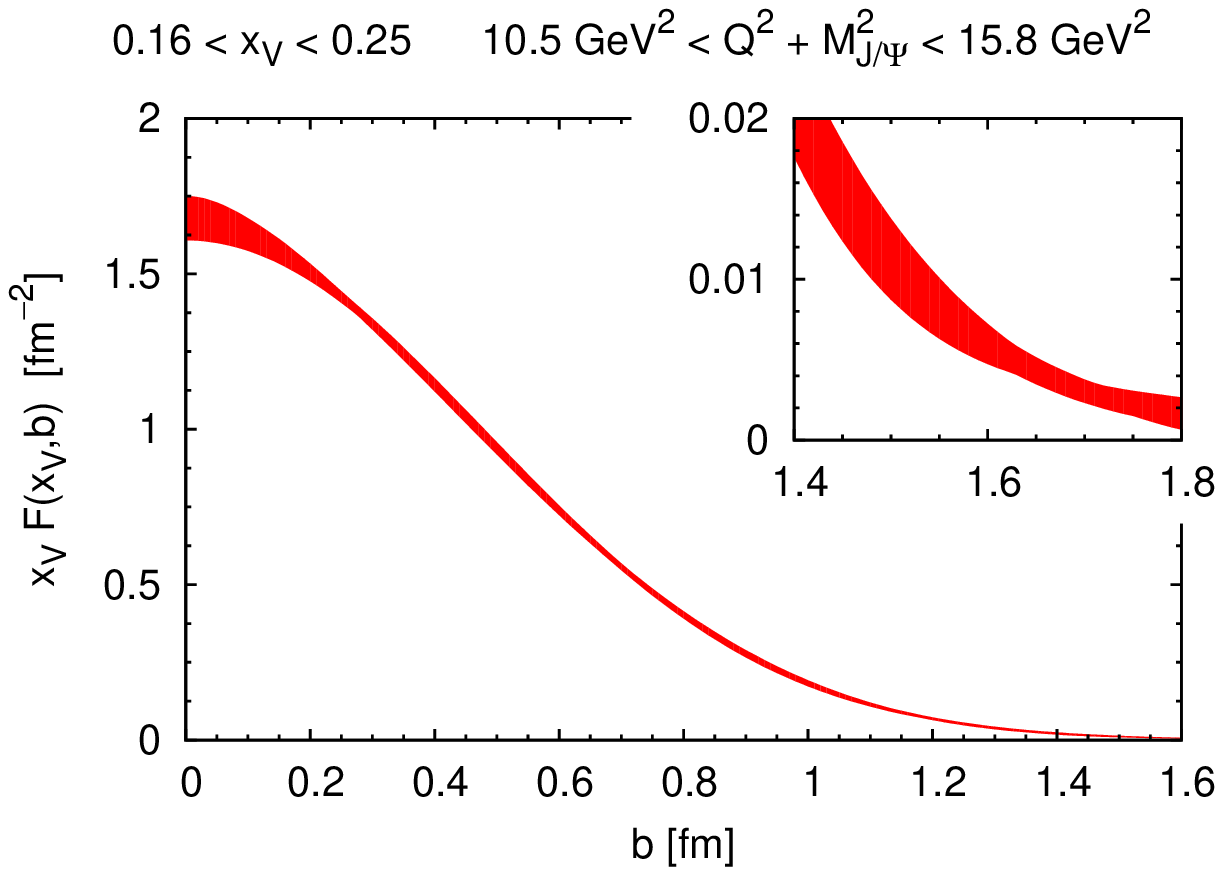}
  \caption{Simulated cross sections for $\gamma^* p\to \jpsi\, p$ and the
    corresponding impact parameter distributions for gluons.  The left
    panels are for $E_e = 20 \operatorname{GeV}$, $E_p = 250
    \operatorname{GeV}$ and the right panels for $E_e = 5
    \operatorname{GeV}$, $E_p = 100 \operatorname{GeV}$, with $10
    \operatorname{fb}^{-1}$ integrated luminosity in both cases.}
  \label{Fig:jpsi}
\end{figure}

Selective information about gluons can be obtained in exclusive $\jpsi$
production.  We have generated pseudo-data for $e^- p\to e^- \jpsi\, p\to
e^- (e^+e^-)\, p$ with a version of {PYTHIA} modified to describe the H1
and ZEUS measurements of this process.  We assume an acceptance in
pseudorapidity of $\eta < 5$ for all final-state leptons and an acceptance
for the recoil proton as in the DVCS studies above.  The $t$ spectra and
corresponding impact parameter distributions in Fig.~\ref{Fig:jpsi} show
that EIC can provide accurate images of gluons in the proton over two
orders of magnitude in $x$.

\section*{Acknowledgments}

I gratefully acknowledge collaboration with E.-C. Aschenauer, S. Fazio,
K. Kumer\v{c}ki and D. M\"uller, who performed the simulations and fits on
which the result presented here are based.


{\raggedright
\begin{footnotesize}

\end{footnotesize}
}

\end{document}